\documentclass[12pt]{article}

\usepackage{epsfig}
\usepackage{amsmath,amssymb,amsfonts}
\usepackage{float,a4wide}
\newcommand{\be}{\begin{equation}}
\newcommand{\en}{\end{equation}}
\newcommand{\bea}{\begin{eqnarray}}
\newcommand{\ena}{\end{eqnarray}}

\newcommand{\hbo}{\hbox to 1 true cm {\hfill } }

\newcommand{\E}{e}

\newcommand{\Tr}{\text{Tr}}
\newcommand{\re}[1]{~(\ref{#1})}

\def\dslash{\partial\kern-.5em\slash}
\def\kslash{k\kern-.5em\slash}
\def\pslash{p\kern-.5em\slash}
\def\Dslash{D\kern-.5em\slash}

\begin{document}
%
%\vglue 1truecm

\vbox{
\hfill CERN-TH/2003-060
}
\vbox{
\hfill UNITU-THEP-05/03
}
\vbox{
\hfill HD-THEP-03-16
}

\vfil
\centerline{\LARGE\bf Casimir Effect on the Worldline}

\vspace{0.5cm}

\bigskip
\centerline{\large Holger Gies${}^{b,c}$, Kurt Langfeld${}^{a}$, Laurent
	Moyaerts${}^{a}$ } 
\bigskip
\vspace{.5 true cm}
{\small
\centerline{ ${}^a$ Institut f\"ur Theoretische Physik, Universit\"at
   T\"ubingen }
\centerline{D-72076 T\"ubingen, Germany}
\bigskip

\centerline{ ${}^{b}$ CERN, Theory Division, CH-1211 Geneva 23, Switzerland }
\bigskip

\centerline{ ${}^{c}$ Institut f\"ur theoretische Physik, Universit\"at
   Heidelberg }
\centerline{D-69120 Heidelberg, Germany}
}
\vspace{1.5 true cm}
\centerline{ March 2003}
\vspace{.5 true cm}

\vskip 1.5cm

\begin{abstract}
\noindent
We develop a method to compute the Casimir effect for arbitrary
geometries. The method is based on the string-inspired worldline
approach to quantum field theory and its numerical realization with
Monte-Carlo techniques. Concentrating on Casimir forces between rigid
bodies induced by a fluctuating scalar field, we test our method with
the parallel-plate configuration. For the experimentally relevant
sphere-plate configuration, we study curvature effects
quantitatively and perform a comparison with the ``proximity force
approximation'', which is the standard approximation technique. Sizable
curvature effects are found for a distance-to-curvature-radius ratio of
$a/R\gtrsim0.02$. Our method is embedded in renormalizable quantum field
theory with a controlled treatment of the UV divergencies. As a
technical by-product, we develop various efficient algorithms for
generating closed-loop ensembles with Gau\ss ian distribution. 

\end{abstract}

\vfil
%\hrule width 5truecm
%\vskip .2truecm
%\begin{quote}
%PACS:
%\end{quote}
\eject

\section{Introduction}
\label{intro}

The Casimir effect \cite{Casimir:dh} has recently been under intense
study, experimentally \cite{Lamoreaux:1996wh} as well as
theoretically (for recent comprehensive reviews, see \cite{review}). In
fact, we are currently witnessing a transition of the Casimir effect
from a pure fundamental quantum effect, being interesting in its own
right, via an experimentally challenging problem to a phenomenon
becoming relevant to applied physics such as nanotechnology
\cite{nano}. Moreover, the Casimir effect has been suggested as an
experimentally powerful tool for investigating new physics beyond the
standard model \cite{Krause:1999ry}.

Considerable progress has been made in recent years as far as the
Casimir effect of real (rather than idealized) conductors is
concerned: the effects of finite conductivity, finite temperature, and
surface roughness are theoretically well under control for the current
experimental realizations. Even the dependence of the Casimir force on
the isotopic composition of the interacting bodies has been studied
recently \cite{Krause:2002wn}. By contrast, the dependence of the
Casimir force on the geometry of the interacting bodies is neither
completely understood nor quantitatively satisfactorily under
control. Except for a small number of analytically solvable
geometries, one has to rely on approximations among which the
``proximity force approximation'' \cite{pft1,pft2} represents the most
widely used method. Roughly speaking, the proximity force
approximation maps the Casimir effect of an arbitrary geometry onto
Casimir's parallel-plate configuration, thereby neglecting curvature
and tilt effects in an uncontrolled manner. In fact, the current
limitations for a quantitative comparison of theory and experiment
arise essentially from an estimated 1\% error of the proximity force
approximation.

The basic obstacles against improving this situation are mainly
technical in nature and partly fundamental. Standard strategies
perform the Casimir calculations in two steps: first, the mode
spectrum of quantum fluctuations in a given background geometry has to
be identified; secondly, the Casimir energy is obtained by summing up
(tracing over) the spectrum. The first step is obviously increasingly
difficult the more complex a given geometry is; without a high degree
of symmetry, even the use of standard numerical techniques is rather
limited. The second step suffers from the same problems, but is
moreover complicated by the fact that the mode sum is generally
ultraviolet divergent. The divergencies have to be analyzed and, if
possible, be removed by renormalization of physical parameters. Not
only is the handling of these divergencies technically (and
numerically) challenging, but the classification of divergencies 
is also still under intense debate
\cite{Deutsch:sc,Graham:2002xq,Milton:2002vm}.

In this work, we propose a method that has the potential to solve
these technical problems. Moreover, it is embedded in perturbative
quantum field theory with its clear and unambiguous renormalization
program. Our method is based on the ``string-inspired'' worldline
formalism in which perturbative $N$-point amplitudes are mapped onto
quantum mechanical path integrals over closed worldlines
\cite{feynman1} (for a recent review, see \cite{Schubert:2001he}). The
technical advantages arise from the fact that the mode spectrum and
its sum are not computed separately but all at once. These worldline
integrals can conveniently be calculated with Monte-Carlo methods
({\em worldline numerics}) with an algorithm that is completely
independent of the Casimir geometry; in particular, no background
symmetry is required. Whereas the worldline integral is 
finite, the ultraviolet divergencies occur in a ``propertime''
integral, roughly corresponding to an integral over the size of the
worldlines. The divergencies can be found at small propertimes
($\hat{=}$ small size $\hat{=}$ ultraviolet), where a mapping to
Feynman-diagram language is possible and the standard rules of
renormalization can be applied.

In order to illustrate our method, we focus in this work on the
calculation of Casimir forces between rigid bodies, induced by quantum
fluctuations of a scalar field. The rigid bodies are modeled by
background potentials $V(x)$ (mainly of $\delta$ function type), which
allow us to approach the idealized limit of Dirichlet boundary
conditions in a controlled way. As a benchmark test, we study the
classic parallel-plate configuration in detail. Finally, we compute
the Casimir forces between a plate and a cylinder as well as the
experimentally highly relevant case of a plate and a sphere, both in
the idealized Dirichlet limit. Here we find clear signals of curvature
effects if the distance between the bodies is roughly a few percent of
the cylinder/sphere radius or larger. This scale characterizes the
limit of quantitative accuracy of the proximity force approximation.

We developed the technique of worldline numerics in~\cite{Gies:2001zp} 
and it has successfully been applied to the
computation of quantum energies or actions induced by scalar or
fermion fluctuations in electromagnetic backgrounds
\cite{Gies:2001zp,Gies:2001tj,Langfeld:2002vy}. As for any numerical
method, possible finite-size or discretization errors have to be
analyzed carefully. In this respect, the idealized Casimir problem
turns out to be most challenging, because the background potentials
with their $\delta$-like support affect the quantum fields on all
scales. Therefore, we have to make sure that our worldline numerics
operates sufficiently close to the ``continuum limit'' (propertime
continuum in our case). We dedicate a whole section
(Sect.~\ref{numerics}) to this question, also relevant for further
applications of worldline numerics, and present a number of new and
efficient algorithms for the generation of Gau\ss ian distributed
closed-loop ensembles.

Though the heart of our method is intrinsically numerical, we would
like to emphasize that the worldline technique offers an intuitive
approach to quantum phenomena. Particularly for Casimir forces
between rigid bodies, many features such as the sign of the
interaction or curvature effects can easily be understood when
thinking in terms of worldline ensembles ({\em loop clouds}). 

The paper is organized as follows: the next section provides a brief
introduction into the worldline approach to the Casimir effect.
Section \ref{numerics} describes efficient methods for the generation
of loop ensembles. The reader who is mainly interested in Casimir
phenomenology may skip this section.  Section \ref{config} provides
for an intuitive understanding of rigid-body Casimir forces in the
light of the worldline language. Our numerical findings for the
rigid-body Casimir force for several geometries (plate-plate,
plate-sphere, plate-cylinder) are presented in section \ref{numres}.

\section{Worldline techniques for Casimir configurations}
\label{worldform}
\subsection{Framework}
\label{setting}

Let us discuss the formalism for the simplest case of a real scalar
field $\phi$ coupled to a background potential $V(x)$ by which we
describe the Casimir configuration. The field theoretic Lagrangian is
\begin{equation}
{\cal L}=\frac{1}{2} \partial_\mu \phi \partial_\mu \phi + \frac{1}{2}
m^2 \phi^2 +\frac{1}{2} V(x)\, \phi^2. \label{1.1}
\end{equation}
The potential $V(x)$ can be considered as a spacetime dependent mass
squared, implying that it has mass dimension 2. In the absence of any
further fields and couplings, the complete unrenormalized quantum
effective action for $V$ is
\begin{eqnarray}
\Gamma[V]&=& \frac{1}{2} \Tr \ln \frac{-\partial^2
  +m^2+V(x)}{-\partial^2 +m^2} \label{1.2fl}\\
&=& -\frac{1}{2}\int_{1/\Lambda^2}^\infty \frac{dT}{T}\int d^D x\, 
    \left[ \langle x| \E^{-T (-\partial^2 +m^2+V(x))}|x\rangle
  -\frac{1}{(4\pi T)^{D/2}} \E^{-m^2T} \right]. \label{1.2}
\end{eqnarray}
Here we work in $D=d+1$ Euclidean spacetime dimensions, i.e., $d$ space
dimensions. In Eq.\re{1.2}, we have introduced the
propertime representation of the $\Tr \ln$ with UV cutoff $\Lambda$
at the lower bound of the $T$ integral.\footnote{Other regularization
techniques are possible as well, e.g., dimensional regularization,
$(dT/T) \to \mu^{2\epsilon}(dT/T^{1-\epsilon})$; the propertime
cutoff is used only for the sake of definiteness. For a pedagogical
review of various regularization techniques in the Casimir
context, see \cite{ReuterDitt}.} Interpreting the
matrix element as a quantum mechanical transition amplitude in
propertime $T$, we can introduce the Feynman path integral, or worldline,
representation,
\begin{equation}
\int d^D x\, 
    \langle x| \E^{-T (-\partial^2 +V(x))}|x\rangle
= \int d^D x_{\text{CM}}\,\,\,\mathcal{N} \!\!\!\!\!\!\!\!
  \int\limits_{x(0)=x(T)}\!\!\!\!
    \mathcal{D} x\, \E^{-\int_0^T  d\tau\, \dot{x}^2/4 -\int_0^T
    d\tau\, V(x_{\text{CM}}+x(\tau))}.\label{transampl}
\end{equation}
The $\Tr$ operation of Eq.\re{1.2fl}, which has led to a transition
amplitude at {\em coincident} points in Eq.\re{1.2}, induces a path
integral over {\em closed} worldlines, $x(0)=x(T)$. In
Eq.\re{transampl}, we have shifted all worldline loops under the
spacetime integral to have a common center of mass $x_{\text{CM}}$,
implying $\int_0^T d\tau\, x_\mu(\tau)=0$. The normalization
$\mathcal{N}$ is determined from the limit of zero potential,
\begin{equation}
 \langle x| \E^{T \partial^2 }|x\rangle \equiv \frac{1}{(4\pi
T)^{D/2}} =\mathcal{N} \!\!\!\!\!\!\!\!
  \int\limits_{x(0)=x(T)}\!\!\!\!
    \mathcal{D} x\, \E^{-\int_0^T  d\tau\, \dot{x}^2/4 },\label{detnorm}
\end{equation}
such that the path integral can be interpreted as an expectation
value with respect to an ensemble of worldlines with Gau\ss ian
velocity distribution,
\begin{equation}
\mathcal{N} \!\!\!\!\!\!\!\!
  \int\limits_{x(0)=x(T)}\!\!\!\!
    \mathcal{D} x\, \E^{-\int_0^T  d\tau\, \dot{x}^2/4 -\int_0^T
    d\tau\, V(x_{\text{CM}}+x(\tau))}
=\frac{1}{(4\pi T)^{D/2}}\, \left\langle \E^{-\int_0^T
    d\tau\, V(x_{\text{CM}}+x(\tau))} \right\rangle_x.
\end{equation}
The last step of our construction that is crucial for numerical
efficiency consists of introducing {\em unit loops} $y(t)$ which are
dimensionless closed worldlines parameterized by a unit propertime
$t\in[0,1]$, 
\begin{equation}
y_\mu(t)=\frac{1}{\sqrt{T}} x_\mu(T t)\quad  \Longrightarrow\quad
\int_0^T d\tau\, \dot{x}^2(\tau) = \int_0^1 dt\, \dot{y}^2(t),
\label{unitloop} 
\end{equation}
where the dot always denotes differentiation with respect to the
argument. 

Inserting the path integral representation\re{transampl} into the
effective action\re{1.2} and using the unit loops $y(t)$, we end up
with the desired formula which is suitable for a numerical
realization,
\begin{equation}
\Gamma[V]=-\frac{1}{2} \frac{1}{(4\pi)^{D/2}} \int_{1/\Lambda^2}^\infty
\frac{dT}{T^{1+D/2}} \, \E^{-m^2T}\int d^D x 
\left[\Big\langle \,W_V[y(t);x]\, \Big\rangle_y -1\right]. \label{1.3}
\end{equation}
Here and in the following we have dropped the subscript ``CM'' of the
center-of-mass coordinate $x_\mu$ and introduced the ``Wilson loop''
\begin{equation}
W_V[y(t);x]=\exp\left[-T \int_0^1 dt\, V(x+\sqrt{T}  y(t))\right],
\label{1.3a} 
\end{equation}
and
\begin{equation}
\Big\langle W_V[y(t);x]\Big\rangle_y = \frac{
\int\limits_{y(0)=y(1)} {\cal D}y\,\, W_V[y(t);x] \,\E^{-\int_0^1 dt\,
  \dot{y}^2/4}}
{\int\limits_{y(0)=y(1)} {\cal D}y\, \,\E^{-\int_0^1 dt\,
  \dot{y}^2/4}} \label{1.4}
\end{equation}
denotes the expectation value of an operator with respect to the path
integral over unit loops $y(t)$. This construction of Eq.\re{1.3} is
exact and completely analogous to the one proposed in
\cite{Gies:2001zp} for electromagnetic backgrounds; further details
can be found therein.

For
time-independent Casimir configurations, we can carry out the time
integration trivially, $\int dx_0=L_{x_0}$, where $L_{x_0}$ denotes the
``volume'' in time direction, and define the (unrenormalized) Casimir
energy as
\begin{equation}
{\cal E}=\Gamma/L_{x_0}. \label{1.5}
\end{equation}
\subsection{Renormalization}
\label{renorm}

The analysis of divergencies in Casimir calculations is by no means
trivial, as the ongoing debate in the literature demonstrates
\cite{Graham:2002xq,Milton:2002vm}. The reason is that divergencies in
these problems can have different sources with different physical
meaning. On the one hand, there are the standard field theoretic UV
divergencies that can be mapped onto divergencies in a finite number
of Feynman diagrams at a given loop order; only these divergencies can
be removed by field theoretic renormalization, which is the subject of
the present section.

On the other hand, divergencies can arise from the modeling
of the Casimir boundary conditions. In particular, idealized
conditions such as perfectly conducting surfaces affect quantum
fluctuations of arbitrarily high frequency; therefore, an infinite
amount of energy may be required to constrain a fluctuating field on
all scales. These divergencies are real and imply that idealized
conditions can be ill-defined in a strict sense. The physically important
question is whether these divergencies affect the physical
observable under consideration (such as Casimir forces) or not. If
not, the idealized boundary conditions represent a simplifying and
valid assumption, and the removal of these divergencies can be
justified. But if the observable is affected, the idealized
conditions have to be dropped, signaling the strong dependence of the
result on the physical details of the boundary conditions (e.g., material
properties). 

Even though the worldline is an appropriate tool for analyzing both
types of divergencies, we concentrate on the first type in this paper,
leaving a discussion of the second for future work. 

In order to isolate the field theoretic UV divergencies, we can expand
the propertime integrand for small propertimes (high momentum scales).
Since this is equivalent to a local gradient expansion in terms of the
potential $V(x)$ (heat-kernel expansion), each term $\sim V(x)^n$
corresponds to a scalar one-loop Feynman diagram with $n$ external
legs coupling to the potential $V(x)$ and its derivatives, and with
the momentum integration already performed (thanks to the worldline
method). Using $\int_0^1 dt\, y_\mu(t)=0$ and $\int_0^1dt\, \langle
y_\mu(t) y_\nu(t)\rangle_y =(1/6) \delta_{\mu\nu}$, we find up to
order $T^2$,
\begin{eqnarray}
\int_x\langle W_V-1\rangle_y
%\equiv\int d^D x \left[\left\langle \E^{-T
%   \int_0^1 dt\, V(x+\sqrt{T}  y(t))}\right\rangle_y -1\right]
&=& - T\int d^Dx\, V(x)
     -\frac{T^2}{6} \int d^Dx\, \partial^2 V(x) \nonumber\\
&&     +\frac{T^2}{2} \int d^Dx\, V(x)^2 +{\cal O}(T^3), \label{ren1}
\end{eqnarray}
which should be read together with the propertime factor $1/T^{1+D/2}$
in Eq.\re{1.3}.  The term $\sim V(x)$ corresponds to the tadpole
graph. In the conventional ``no-tadpole'' renormalization scheme, the
renormalization counter term $ \sim V(x)$ is chosen such that it
cancels the tadpole contribution completely. Of course, any other
renormalization scheme can be used as well. The corresponding counter
term can be fixed unambiguously by an analysis of the tadpole Feynman
diagram in the regularization at hand. In $D<4$ spacetime dimensions,
there is no further counter term, since $V(x)$ has mass dimension
2. The remaining terms of ${\cal O}(T^2)$ are UV finite in the limit
$T\to 1/\Lambda^2\to 0$.

In $4\leq D<6$, we need further subtractions. Here, it is useful to
note that the last term on the first line of Eq.\re{ren1} vanishes
anyway, provided that the potential is localized or drops off
sufficiently fast at infinity. This is, of course, always the case for
physical Casimir configurations.\footnote{Strictly speaking,
infinitely extended surfaces such as idealized infinitely large plates
do not belong to this class, but we can always think of large but
finite surfaces and then take the infinite-surface limit {\em after}
the infinite-volume limit.} Renormalization provides us with a further
counter term $\sim \int_x V^2$ subject to a physically chosen
renormalization condition such that the divergence arising from the
last $T^2$ term is canceled. With this renormalization condition, the
physical value of the renormalized operator $\sim V^2$ is
fixed.\footnote{Since we used a gradient expansion, the renormalized
operator is fixed in the small-momentum limit; if the renormalization
condition operates at finite momentum, e.g., using the polarization
operator, possible finite renormalization shifts can be obtained from
an analysis of the corresponding Feynman diagram. However, in the
present case of static Casimir problems, it is natural to impose a
renormalization condition in the small-momentum limit anyway.} For
even higher dimensions, similar subtractions are required that involve
higher-order terms not displayed in Eq.\re{ren1}.

As far as controlling divergencies by renormalization is concerned,
this is all there is and no further {\em ad hoc} subtractions are
permitted. However, having removed these UV divergencies with the
appropriate counter terms does not guarantee that the resulting Casimir
energy is finite. Further divergencies may arise from the form of the
potential as is the case for the idealized Casimir energies 
mentioned above. 

In the present work, we take up a more practical position and are
merely interested in the Casimir forces between disconnected rigid bodies
which are represented by the potential $V(x)=V_1(x)+V_2(x)+\dots$. We
assume the rigid bodies as given, disregarding the problem of whether
the Casimir energy of every single body is well defined by itself. For
this, it suffices to study the {\em interaction} Casimir energy defined as
the Casimir energy of the whole system minus the separate energies of
the single components,
\begin{equation}
E:={\cal E}_{V=V_1+V_2+\dots}-{\cal E}_{V_1} -{\cal E}_{V_2} -\dots 
\; \; . 
\label{ren2}
\end{equation}
Note that the subtractions do not contribute to the Casimir force
which is obtained by differentiating the interaction energy with
respect to parameters that characterize the separation and orientation
of the bodies.  By this differentiation, the subtractions drop
out. Furthermore, these terms remove the field theoretic UV
divergencies of Eq.\re{ren1}: this is obvious for the terms linear in
$V(x)$; for the quadratic one, this follows from $\int_x
V^2=\int_x(V_1+V_2+\dots)^2=\int_x(V_1^2+V_2^2+\dots)$. The last
equation holds because of the local support of the disconnected
bodies. By the same argument, the subtractions remove every term of a
local expansion of ${\cal E}_{V_1+V_2+\dots}$ to any finite order. In
this way, any divergence induced locally by the potentials is
canceled. But, of course, the Casimir force is not removed -- it is
inherently nonlocal.

The interaction energy in Eq.\re{ren2} is also numerically favorable,
since the subtractions can be carried out already on the level of the
propertime integrands, avoiding manipulations with large numbers. 

We would like to stress that the definition of the interaction energy
in Eq.\re{ren2} should not be confused with renormalization. It is a
procedure for extracting exact information about the Casimir force
between rigid bodies, circumventing the tedious question as to whether
Casimir energy densities are locally well defined. This procedure
also removes the field theoretic UV divergencies. In this case,
renormalization conditions which fix the counter terms do not have to
be specified. These local counter terms cannot exert an influence on
the Casimir force for disconnected rigid bodies anyway, because
the latter is a nonlocal phenomenon. Expressed in physical terms of
the QED Casimir effect: the renormalized strength of the coupling
between the electromagnetic field and the electrons in the metal is,
of course, important for a computation of the local energy density
near a plate, but the Casimir force between two plates is independent
of the electromagnetic coupling constant.

We would like to point out that the concept of the interaction energy is
meaningless for the computation of Casimir stresses of single bodies,
e.g., a sphere. Here, the renormalization procedure has to be carried
out as described above, and the result may depend on the
renormalization conditions and strongly on the details of the
potential.

\section{Worldline numerics}
\label{numerics}

In this section, we discuss possible numerical realizations of the
worldline integral Eqs.\re{1.3}-\re{1.4} (the more phenomenologically
interested reader may proceed directly to Sect.~\ref{config}). 

As proposed in~\cite{Gies:2001zp}, we estimate the analytical integral
over infinitely many closed worldlines by an ensemble average over
finitely many closed loops obeying a Gau\ss ian velocity distribution
$P[\{y(t)\}]$,
\begin{equation}
P[\{ y(t)\}]=\delta\left( \int_0^1 dt\, y(t) \right) \, \exp \left(
-\frac{1}{4} \int_0^1 dt\, \dot{y}^2\right), \quad \text{with}\,\,
y(0)=y(1), \label{wn1}
\end{equation}
where the $\delta$ constraint ensures that the loops are centered upon
a common center of mass (here and in the following, we drop all
normalizations of the distributions, because they are irrelevant when
taking expectation values). Here, we have chosen to work with rescaled
{\em unit loops} $y(t)$ as introduced in
Eqs.\re{1.3}-(\ref{1.4}). Numerical arithmetics requires discretization;
however, we generally {\em do not} discretize spacetime on a lattice,
but only the loop propertime parameter $t$:
\begin{equation}
\{y(t)\}\quad\to\quad \{y_k\}\in \mathbb{R}^D, \quad k=1,2,\dots,N,
 \label{wn2}
\end{equation}
where $N$ denotes the number of points per loop (ppl).  Whereas
Gau\ss ian distributed numbers can easily be generated, the numerical
difficulty is to impose the $\delta$ constraint,
$y_1+y_2+\dots+y_N=0$, and the requirement of closeness. In the
following, we discuss four possible algorithms, and recommend the last
two of them based on Fourier decomposition (``f loops'') or
explicit diagonalization (``v loops'').

\subsection{ Heat-bath algorithm} 

A standard approach for the generation of field (or path)
distributions that obey a certain action is the heat-bath algorithm,
which has been employed for worldline numerics in
\cite{Gies:2001zp,Gies:2001tj,Langfeld:2002vy}. Discretizing the
derivative in the exponent of Eq.\re{wn1}, e.g., by $\dot y \to N (
y_k-y_{k-1})$, each point on a loop can be regarded as exposed to a
``heat bath'' of all neighboring points. The discretized probability
distribution then reads
\begin{equation}
P\bigl[ \{ y_k \} \bigr] \; = \; 
\delta \Bigl( y_1 + \ldots + y_N \Bigr) \; 
\exp \Bigr\{ - \frac{N}{4} \sum _{k=1}^N (y_{k} - y_{k-1})^2
\Bigr\} \; . 
\label{eq:a1} 
\end{equation} 
where $y_0\equiv y_N$. The heat-bath procedure now consists in the
following steps: (i) choose a site $i\in [1,N]$, consider all
variables $y_k$, $k\neq i$ as constant, and generate the $y_i$
according to its probability; (ii) visit all variables of the loops
(e.g., in a serial fashion or using the checkerboard algorithm).
Thereby, the closeness requirement is easily realized with, e.g.,
$y_N$ being in the heat bath of $y_{N-1}$ and $y_1$, etc. The
center-of-mass constraint can be accommodated by shifting the whole
loop correspondingly after one thermalization sweep (update of all
points per loop).

Whereas this procedure has been sufficient for the applications
discussed in \cite{Gies:2001zp,Gies:2001tj,Langfeld:2002vy}, it turns
out that this algorithm suffers in practice from a thermalization
problem for large values $N$.  To demonstrate this, let us define the
extension $e$ of the loop ensemble by the loop mean square
\begin{equation} 
e^2 \; = \; \frac{1}{\cal N} \int dy_1 \ldots dy_N \; y_k^2 \; 
P\bigl[ \{ y_k \} \bigr] \; , \hbo 
{\cal N} \; = \; \int dy_1 \ldots dy_N \; 
P\bigl[ \{ y_k \} \bigr] \; . 
\label{eq:a2} 
\end{equation} 
This quantity can be calculated analytically, straightforwardly
yielding  
\begin{equation}
e \; = \; \sqrt{ \frac{1}{6}{ \left( 1 - \frac{1}{N^2} \right) }} \; . 
\label{eq:a3} 
\end{equation}
In order to generate a ``thermalized'' loop, one starts with a random
ensemble $\{y_k\}$ and performs $n_t$ heat-bath sweeps. For each loop,
we calculate its extension $e$. After averaging over 1000 loops,
we compare the estimator of $e$ as function of $n_t$ with the analytic
result (\ref{eq:a3}) corresponding to the limit $n_t\rightarrow
\infty$.

\begin{figure}[t]
\centerline{  
\epsfxsize=8cm 
\epsffile{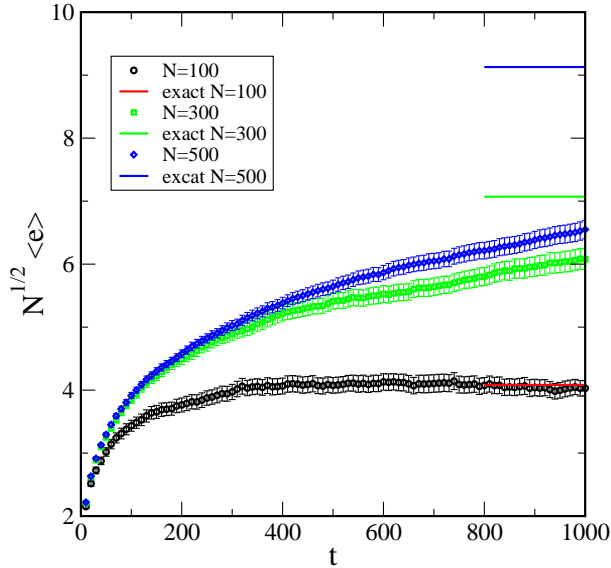} 
} 
\caption{ The average extension $e$ (multiplied by $\sqrt{N}$ for better
visualization) of the loops as function of the 
   number of thermalizations $n_t$. }
\label{fig:1}
\end{figure}

The result is shown in Fig.~\ref{fig:1}.  One clearly observes that
the thermalization of loop ensembles is expensive for $N>500$. In fact,
roughly $n_t=45000$ is needed for an acceptable loop ensemble
consisting of $N=1000$ points. Since a computation of Casimir energies
requires loop ensembles of $N\geq 1000$, the heat-bath algorithm is
too inefficient and cannot be recommended.

\subsection{Random Walk}

In order to circumvent the thermalization problem, one may exploit the
connection between loops with Gau\ss ian velocity distribution and
random walks \cite{Itzykson:sx,Samuel}. This has been adapted to
worldline numerics with latticized spacetime in
\cite{Schmidt:2002yd}; here, however, we keep spacetime continuous. 
For this purpose, let us give up the concept of unit loops for a
moment, and reinstate the naturally emerging coordinate space loops
$x(\tau)$,
\begin{equation}
x(\tau)=\frac{1}{\sqrt{T}}\, y(\tau/T), \quad x(0)=x(T). \label{wn4}
\end{equation}

Probability theory tells us that random walks automatically implement
the Gau\ss ian velocity distribution 
\begin{equation} 
\prod_{i=1}^{N-1} \exp \biggl\{ 
-\frac{1}{4\Delta\tau} (x_{i+1}-x_i)^2 \biggr\}. \label{gm} 
\end{equation}
The crucial point is to establish the relation between a loop that a
random walker with step length $s$ would generate for us and a
thermalized loop at a given propertime $T$. This relation results from
a coarse-graining procedure, which we present here briefly.  Given
that the random walker starts at the point $x_i$, the probability
density for reaching the point $x_f$ after $n$ steps is given by
$$ 
p(x_f \mid x_i,n,s) = \int d^Dx_2 \dots d^Dx_{n-1}\,\,
\prod_{k=1}^{n-1} \frac{1}{\Omega(D) s^{D-1}}
\delta(\mid\!x_{k+1}-x_{k}\!\mid -s), 
$$
with $\Omega(D)$ being the solid angle in $D$ dimensions, $x_1=x_i$
and $x_n=x_f$.  For $n \gg 1$, but $ns^2$ fixed, the central-limit
theorem can be applied~\cite{Samuel}:
\begin{equation}
\lim _{n \to \infty }  p(x_f \mid x_i,n,s) \; = \; 
\left(\frac{D}{2 \pi n s^2}\right)^{\frac{D}{2}} 
\exp \biggl\{-\frac{D}{2
ns^2}\left(x_f-x_i\right)^2  \biggr\}, \; \; \; \; 
ns^2 = \mathrm{fixed}  . 
\label{clt} 
\end{equation}
Comparing (\ref{clt}) with (\ref{gm}), one identifies 
\begin{equation}
\Delta\tau=\frac{n s^2}{2D}. 
\end{equation}
The dimension of the propertime as well as its relation to the loop
length $L$ appears here in an obvious way,
\begin{equation}
T=\frac{N_w s^2}{2D}=\frac{L s}{2D}, \label{tl}
\end{equation}
where $N_w$ now is the total number of walker steps.  Is is important
to point out that the propertime can be tuned in two ways: we can
adjust either the walker step $s$ or the number of points $N_w$.  The
corresponding two methods to generate a loop ensemble at given
propertime $T$ work as follows: \vspace{0.3cm} \\ \underline{Method 1}
: $s$ is fixed.
\begin{itemize}
\setlength{\itemsep}{-0.7mm} 
\item[(1)] choose the walker step $s$; 
\item[(2)] read off from Eq.~(\ref{tl}) the number of points $N_w$
corresponding to $T$; 
\item[(3a)] generate $N_w$ points by letting a random walker go $N_w$
steps, and accept the configuration if the last step leads him into a
small sphere (radius $\varepsilon $) centered upon the starting point; 
\item[(3b)] close the loop 'by hand' by shifting the last point to the 
starting point; 
\item[(4)] shift the center of mass to zero; 
\item[(5)] repeat steps (3) and (4) $n_{\text{L}}$ times for an ensemble
of $n_{\text{L}}$ loops. 
\end{itemize}
We point out that the value of $s$ must be much smaller than 
the characteristic length scale provided by the background potential. 
A second constraint on $s$ arises from the applicability of the
central limit theorem, i.e., $n\gg1$ in (\ref{clt}).  
A third systematic numerical uncertainty follows from the shift 
in step (3b). Unfortunately, small values for $\varepsilon $ 
result in low acceptance rate for loops, and, therefore, increase the 
numerical effort to generate the loop ensemble. A
good compromise is to set the radius $\varepsilon $ to some percentage 
of the step length $s$.  

\begin{figure}[t]
\centerline{ 
\epsfxsize=10cm 
\epsffile{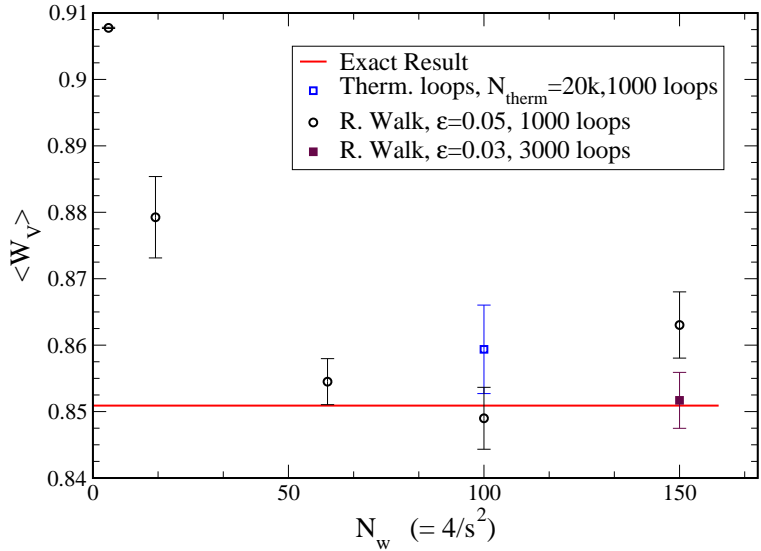} 
} 
\caption{ Average Wilson Loop $\langle W_{V}\rangle$ (cf.
Eq.~(\ref{1.4})) for the case of a constant magnetic background field
$B$ for $B=1$, $T=1$ and $D=2$ as a function of the number of points per
loop. }
\label{fig:2}
\end{figure}

For illustration only, we leave the Casimir effect for a second and
consider the average Wilson Loop $\langle W_V\rangle$ (see
Eq.~(\ref{1.4})) for the case of a constant magnetic background field
$\vec{B}=B\vec{e_z}$ at $T=1$ and $D=2$,
$$ 
V(x) \; = \; A_k (x) \, \dot{x}_k \; , \; \; \; \; \; 
\vec{A}=B/2\,(y,-x,0) \; . 
$$
For $T=1$ the walker step length is given by $s=\frac{2}{\sqrt{N_w}}$.
Figure \ref{fig:2} shows our numerical result as a function of $N_w$ in
comparison with the exact value.  Circles with error bars correspond
to loop ensembles generated with $\varepsilon=0.05 \, s$. The limit
(\ref{clt}) seems to be attained for $50<N_w<100$ ($s<0.3$).  For a
further improvement of the numerical accuracy, large values of $N_w$
and a decrease of $\varepsilon $ at the same time are required.
Finally, we point out that the deviation from the exact result in the
case of the heat-bath-generated loop ensemble (blue square) is
probably due to thermalization effects.

Note that we have to generate a loop ensemble for each value of $T$
($\sim N_w s^2$), which makes this procedure far more memory consuming
than the heat-bath approach.  If we decide to generate the loop
ensembles once and for all and save them to disk, we have to handle
huge amounts of data. On the other hand, if we create our loops 'on
demand' (while performing the $T$ or $\vec{x}$ integrations), we are
confronted with a serious waste of computing time. \vspace{0.3cm} \\
\underline{Method 2} : $N_w$ is fixed.
\begin{itemize}
\setlength{\itemsep}{-0.7mm} 
\item[(1)] choose the number of points $N_w$;
\item[(2)] set the walker step to $s=1$; 
\item[(3)] proceed with steps (3), (4) and (5) of the first method.
\end{itemize}
The loop ensemble is here generated only once and then rescaled to
adjust the step length to the value $s$ corresponding to $T$ in
(\ref{tl}). This method therefore works as in the case of the
rescalable thermalized unit loops, with the difference that the
propertime rescaling is realized via the rescaling of the step length.
This tuning at the level of $s$ provides for a better control of the
microscopic features of the loops.  The second procedure is thus a
good candidate to replace the thermalized loops since it combines the
absence of thermalization and the rescaling of an all-at-once
generated ensemble.

It should however be emphasized that most of the
computer time is spent on generating redundant open loops.  This
is due to the fact that, for a given $\varepsilon $, the fraction of
loops which close after $N_w$ steps decreases like $N_w^{-D/2}$.  

\subsection{Fourier decomposition: ``f loops''} 

We are now looking for alternative methods that could combine some
advantages of the two previous approaches and bypass the problems
rendering them impractical.  A highly efficient procedure
arises from a Fourier representation of our unit loops
\begin{equation}
y(t) \; = \; \sum _{\nu = 0 }^N \Biggl[ a_\nu \; \cos \Bigl( 
2\pi \, \nu \, t \Bigr) \; + \; b_\nu \; \sin \Bigl( 
2\pi \, \nu \, t \Bigr) \Bigr] \; , \hbo a_0 \; = \; 0 \; , 
\label{fo1}
\end{equation}
where $N$ is the number of Fourier modes included (which agrees with the
number of points specifying each loop, see below). The choice
$a_0=0$ guarantees that the loop center of mass is located at the
origin. Inserting Eq.~(\ref{fo1}) into Eq.~(\ref{wn1}), the
probability distribution for the coefficients is given by
\begin{equation}
P\bigl[a,b\bigr] \; = \; \exp \Bigl\{ - \frac{ \pi ^2 }{2} 
\sum _{\nu=1}^N a_\nu^2 \; - \; \frac{ \pi ^2 }{2} \sum _{\nu=0}^N b_\nu^2 
\Bigr\} \; . 
\label{fo2}
\end{equation}

We can then take advantage of the fact that the Fourier components $\{
a,b\}$ are not correlated, in order to generate our loops in momentum
space. The reconstruction of the unit loop $y(t)$ in Eq.~(\ref{fo1}) is
most efficiently performed by using the fast Fourier transformation
(FFT). For these purposes, we define complex coefficients $c_\nu :=
a_\nu + i \, b_\nu $, and obtain
\begin{equation}
y(t) \; = \; \Re \sum _\nu c_\nu \; \exp \Bigl\{ -i \, 2\pi \, t \, \nu 
\Bigr\} \; . 
\label{fo3}
\end{equation}
The FFT procedure generates a series of points $y_i$, $i=0 \ldots N-1$ 
which discretize the continuous curve $y(t)$ and thereby constitute
the unit loops. 

\subsection{Explicit diagonalization: ``v loops''}

Finally, we propose an algorithm that is based on a linear
variable transformation $\{y_k\} \to \{\bar{v}_k\}$, such that the
discretized distribution\re{eq:a1} becomes purely Gau\ss ian. These new
variables are velocity-like and diagonalize the quadratic form in the
exponent. 

Because of the $\delta$ function in Eq.\re{eq:a1}, only $N-1$
coordinates per loop are independent. Defining $\int {\cal
D}y=\int_{-\infty}^\infty \prod\limits_{i=1}^N dy_i$, we may perform,
e.g., the $y_N$ integration using the $\delta$ function,
\begin{eqnarray}
\int{\cal D}y\, P\bigl[ \{ y_k \} \bigr] \dots &=&\int
\prod\limits_{i=1}^{N-1} dy_i\, \, \E^{\left[ -\frac{N}{4} \left(
\sum_{i=2}^{N-1} (y_i-y_{i-1})^2 +(2y_1+y_2+\dots+y_{N-1})^2
+(y_1+y_2+\dots+2 y_{N-1})^2 \right) \right]} \dots \nonumber\\
&=:& \int
\prod\limits_{i=1}^{N-1} dy_i\, \, \E^{\left[ -\frac{N}{4} Y\right]}
\dots, \label{B2}
\end{eqnarray}
where the dots represent an arbitrary $y$-dependent operator, and we
introduced the abbreviation $Y$ for the quadratic form. In order to
turn the exponential into a product of simple Gau\ss ians, we define
$N-1$ new velocity-like variables, 
\begin{eqnarray}
\bar v_1&:=&\frac{3}{2} y_1 +y_2+y_3+\dots+y_{N-2}+\frac{3}{2} y_{N-1},
\nonumber\\
v_i&:=&y_i-y_{i-1}, \quad i=2,3,\dots,N-1. \label{B3}
\end{eqnarray}
For notational simplicity, it is useful to also introduce the
auxiliary variable, 
\begin{equation}
v_{i,j}=v_i+v_{i-1}+\dots+v_{j+1} \equiv y_i-y_j, \quad \text{for}\,\,
i\geq j=1,2,\dots, N-1, \label{B4}
\end{equation}
such that the exponent $Y$ can be written as
\begin{eqnarray}
Y&=&\sum_{i=2}^{N-1} v_i^2 +\left(\bar v_1-\frac{1}{2}
v_{N-1,1}\right)^2
+\left(\bar v_1+\frac{1}{2} v_{N-1,1}\right)^2 \nonumber\\
&=& 2\bar v_1^2 + \frac{1}{2}
v_{N-1,1}^2+ \sum_{i=2}^{N-1} v_i^2 . \label{B5}
\end{eqnarray}
We observe that the variable $\bar v_1$ now appears quadratically in
the exponent as desired. The same has still to be achieved for
$v_2\dots v_{N-1}$. For this, we note that $v_{N-1,1}=v_{N-1}
+v_{N-2,1}$ by definition\re{B4}. Defining 
\begin{equation}
\bar v_{N-1}:=v_{N-1}+\frac{1}{3} v_{N-2,1}, \label{B6}
\end{equation}
we indeed obtain for the exponent $Y$
\begin{eqnarray}
Y&=&2\bar v_1^2 + v_{N-1}^2 + \frac{1}{2}
(v_{N-1} +v_{N-2,1})^2 +\sum_{i=2}^{N-2} v_i^2  \nonumber\\
&=& 2\bar v_1^2 +\frac{3}{2} \bar v_{N-1}^2 +\frac{1}{3} v_{N-2,1}
+\sum_{i=2}^{N-2} v_i^2, \label{B7}
\end{eqnarray}
where $\bar v_{N-1}^2$ also appears quadratically. We can continue
this construction by defining
\begin{equation}
\bar v_{N-i} :=v_{N-i} + \frac{1}{i+2}\, v_{N-i-1,1}, \quad
i=1,\dots,N-2 \; , 
\label{B8} 
\end{equation}
which turns the exponent $Y$ into a purely Gau\ss ian form:
\begin{equation}
Y=2\bar v_1^2 +\frac{3}{2} \bar v_{N-1}^2 +\frac{4}{3} \bar v_{N-2}^2
+\dots +\frac{i+2}{i+1} \bar v_{N-i}^2 +\dots + \frac{N}{N-1} \bar
v_2^2. \label{B9}
\end{equation}
The last step of this construction consists in noting that we can
substitute the integration variables according to 
\begin{equation}
\prod_{i=1}^{N-1} dy_i =J \prod_{i=2}^{N-1} dv_i d\bar v_1
=\bar J \prod_{i=1}^{N-1} d\bar v_i \equiv {\cal D}\bar v \label{B10}
\end{equation}
with nonzero but constant Jacobians $J$, $\bar J$, the value of which
is unimportant for the calculation of expectation values. This allows
us to write the path integral Eq.\re{B2} as
\begin{equation}
\int {\cal D}y\, P\bigl[\{ y_k\} \bigr] \dots 
= \bar J \int {\cal D} \bar v\,\, \exp \left[ -\frac{N}{4} \left( 2
\bar v_1^2 +\sum_{i=1}^{N-2} \frac{i+2}{i+1} \bar v_{N-i}^2 \right)
\right] \dots \equiv \bar J \int {\cal D} \bar v\, P\bigl[\{ \bar
v_k\} \bigr] \dots, \label{B11}
\end{equation}
where $P\bigl[\{ \bar v_k\} \bigr]$ can now be generated
straightforwardly with the Box-M\"uller method \cite{BoxMueller}. 

For the construction of unit loops (``v loops''), the above steps have
to be performed backwards. The recipe is the following:
\begin{itemize}
\setlength{\itemsep}{-0.7mm} 
\item[(1)] generate $N-1$ numbers $w_i$, $i=1,\dots, N-1$ via the
Box-M\"uller method such that they are distributed according to $\exp
(- w_i^2)$; 

\item[(2)] compute the $\bar v_i$, $i=1,\dots,N-1$, by normalizing the
$w_i$:
\begin{eqnarray}
\bar v_1 &=& \sqrt{ \frac{2}{N}}\, \, w_1, \nonumber\\
\bar v_i &=& \frac{2}{\sqrt{N}} \sqrt{ \frac{N+1-i}{N+2-i}}\,\, w_i,
\quad i=2,\dots,N-1 \; ;  \label{B12}
\end{eqnarray}

\item[(3)] compute the $v_i$, $i=2,\dots,N-1$, using
\begin{equation}
v_i=\bar v_i -\frac{1}{N+2-i}\, v_{i-1,1}, \quad \text{where}\,\,
v_{i-1,1} =\sum_{j=2}^{i-1} v_j \; ; 
\label{B13} 
\end{equation}

\item[(4)] construct the unit loops according to 
\begin{eqnarray}
y_1&=& \frac{1}{N} \left( \bar v_1 -\sum_{i=2}^{N-1} \left( N-i+
\frac{1}{2} \right) v_i \right), \nonumber\\
y_i&=& y_{i-1} +v_i, \quad i=2,\dots, N-1, \nonumber\\
y_N&=& -\sum_{i=1}^{N-1} y_i \; ; 
\label{B14}
\end{eqnarray}

\item[(5)] repeat this procedure $n_{\text{L}}$ times for
$n_{\text{L}}$ unit loops.

\end{itemize} 

The formulas in step (4) can be checked straightforwardly by inserting
the definitions of the $v_i$'s and $\bar v_1$. 

This v-loop algorithm allows us to generate unit loops efficiently without
thermalization, i.e., no redundant thermalization sweeps have to be
performed, and works for an arbitrary number of points per loop $N$. 

\subsection{Benchmark test}

We test the quality of our loops with the aid of the Casimir energy
for the parallel-plate configuration in the Dirichlet limit, the
physics of which is described in the next section.

As far as numerics is concerned, there are basically two parameters
that control the quality of our loop ensemble: the number  of
points per loop (ppl) $N$, and the number of loops $n_{\text{L}}$. The
larger these numbers, the more accurate is our numerical estimate at
the expense of CPU time and size. Whereas increasing the number of
loops $n_{\text{L}}$ reduces the statistical error of the Monte-Carlo
procedure, increasing the number of ppl $N$ reduces the systematic
error of loop discretization. 

In order to estimate this systematic error, we have to study the
approach towards the continuum limit. The idea is to choose $N$ large
enough for a given $n_{\text{L}}$, such that the systematic error is
smaller than the statistical one. 

\begin{figure}
\begin{center} 
\epsfig{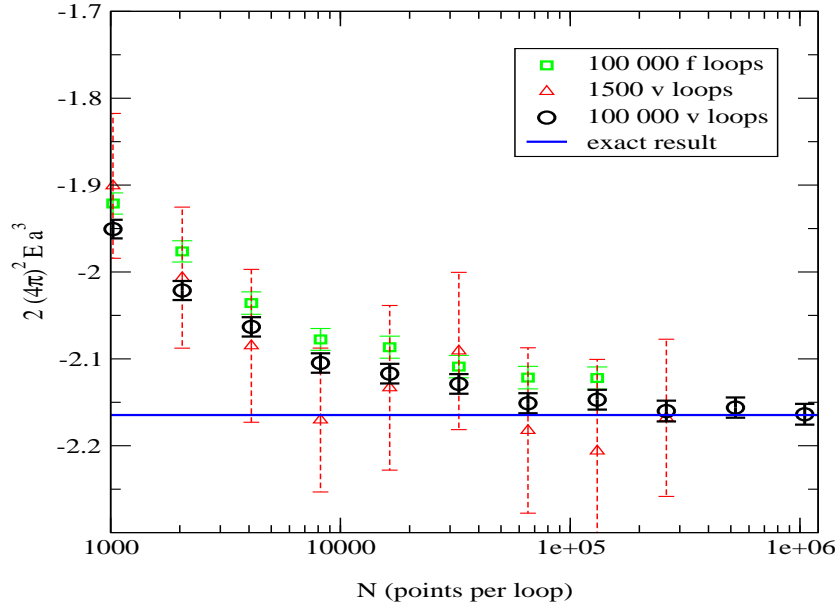}
\end{center}
\caption{Numerical estimate of the interaction Casimir energy of the
parallel-plate configuration for various loop ensembles as a function
of the number of points per loop $N$. The error bars correspond to the
Monte-Carlo statistical error; deviations from the exact
result on top of the statistical error measure the systematic error
due to loop discretization.}
\label{figppl}
\end{figure}

In Fig.~\ref{figppl}, we plot the numerical estimates for the
parallel-plate Casimir energy as a function of the number of ppl $N$
and compare it with the classic result. The error bars represent the
statistical error of the Monte-Carlo procedure. The deviation of the
numerical estimates from the exact result on top of the error bars
serves as a measure of the systematic error. As is visible therein, a
rather small number of several thousand ppl, $N\gtrsim{\cal O}(1000)$,
is sufficient to get a numerical estimate with $\lesssim 5\%$ error
using $n_{\text{L}}=1500$ loops. For a high-precision estimate with an
error $\lesssim 0.5\%$, larger loop ensembles with $n_{\text{L}}\gtrsim
100\,000$ are required. For $N\simeq 50\,000$ppl, systematic and
statistical errors are of the same order, and for $N\gtrsim
100\,000$ppl, the systematic error is no longer relevant for v
loops. For f loops, however, we observe a systematic 1\% error in the
high-precision data of unclear origin.

Nevertheless, the important conclusion of this test is that worldline
numerics has proved its ability to describe quantum fluctuations with
Dirichlet boundary conditions quantitatively.

\section{Casimir forces between rigid bodies}
\label{config}

Casimir forces can be analytically computed for only a small number of
rigid-body geometries among which there is Casimir's classic result
for the parallel-plate configuration; for perfectly conducting plates
at a distance $a$, the interaction energy per unit area is
\cite{Casimir:dh} 
\begin{equation}
E_{\text{PP}}(a)=-\frac{1}{2} \frac{\pi^2}{720} \, \frac{1}{a^3} 
\label{classic}
\end{equation}
for a fluctuating real scalar field; for a complex scalar as well as
for electromagnetic fluctuations, the factor 1/2 has to be
dropped. The famous Casimir force is obtained by differentiating
Eq.\re{classic} by $a$.

\subsection{Proximity force approximation}

The standard approximation method for not analytically solvable
Casimir problems is the proximity force approximation (PFA)
\cite{pft1,pft2}. The basic idea is to apply the parallel-plate result
to infinitesimal bits of the generally curved surfaces and integrate
them up,
\begin{equation}
E=\int_S E_{\text{PP}}(z)\, d\sigma,\label{pft}
\end{equation}
where $E_{\text{PP}}$ is the interaction energy per unit area of the
parallel-plate case. $S$ represents the integration domain and denotes
either one of the surfaces of the interacting bodies or a suitably
chosen mean surface \cite{pft2}. At this point, the proximity force
approximation is ambiguous, and we will simply insert both surfaces in
order to determine the variance. In Eq.\re{pft}, $d\sigma$ denotes the
invariant surface measure, and $z$ represents the separation between
the two surfaces associated with the surface element $d\sigma$ on
$S$. Obviously, the proximity force approximation neglects any
nonparallelity and any curvature -- the latter because each surface
element on $S1$ is assumed to ``see'' only one surface element on $S2$
at separation $z$; but curvature effects require information about a
whole neighborhood of the element on $S2$.

The proximity force approximation is expected to give reasonable
results only if (i) the typical curvature radii of the surfaces
elements is large compared to the element distance and (ii) the
surface elements with strong nonparallelity are further separated than
the more parallel ones.\footnote{The second condition is not so well
discussed in the literature; it is the reason why the proximity force
approximation gives reasonable results for a convex spherical lens
over a plate (convex as seen from the plate), but fails for a concave
lens.}

For configurations that do not meet the validity criteria of the
proximity force approximation, a number of further
approximations or improvements exist, such as an additive summation of
interatomic pairwise interactions and the inclusion of screening
effects of more distant layers by closer ones \cite{review}. Though
these methods have proved useful and even quantitatively precise for a
number of examples, to our knowledge, a general, unambiguous and 
systematically improvable recipe without {\em ad hoc} assumptions is
still missing. 

In Sect.~\ref{numres}, we compare our results with the proximity force
approximation in the simplest version as mentioned above, in order to
gain insight into the effects of curvature. 
 
\subsection{Casimir forces on the worldline}

As described in Sect.~\ref{worldform}, we represent the rigid bodies by a
potential $V(x)$. The functional form of the potential leaves room
enough for modeling many physical properties of real Casimir
configurations. Let us confine ourselves to an idealized potential
well which is represented by a $\delta$ function in space (for ``soft''
boundary conditions, see, e.g., \cite{Actor:vc}),
\begin{equation}
V(x)=\lambda\int_\Sigma d\sigma\, \delta^d(x-x_\sigma), \label{2.1}
\end{equation}
where the geometry of the Casimir configuration is represented by
$\Sigma$, denoting a $d-1$ dimensional surface. $\Sigma$ is generally
disconnected (e.g., two disconnected plates, $\Sigma=S_1+S_2$) and can
be degenerate, i.e., effectively lower dimensional (a point). The
surface measure $d\sigma$ is assumed to be reparametrization
invariant, and $x_\sigma$ denotes a vector pointing onto the
surface. The coupling $\lambda$ has mass dimension 1 and is assumed to
be positive. It can roughly be viewed as a plasma frequency of the
boundary matter: for fluctuations with frequency $\omega\gg\lambda$,
the Casimir boundaries become transparent. In the limit $\lambda\to
\infty$, the potential imposes the {\em Dirichlet boundary condition},
implying that all modes of the field $\phi$ have to vanish on
$\Sigma$.
 
\begin{figure}
\begin{center}
\epsfig{figure=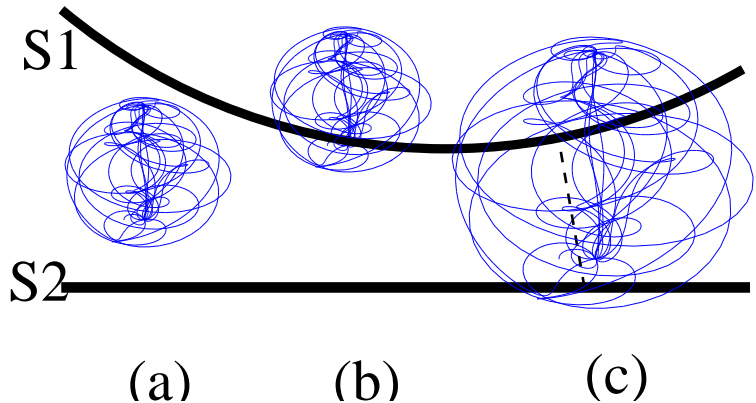,width=10cm}
\end{center}
\caption{Worldline loop contributions to Casimir energies between two
surfaces (S1 and S2): loop (a) does not contribute at all, it is an
ordinary vacuum fluctuation. Loop (b) contributes to the local energy
density near the upper plate, but does not contribute to the Casimir
force. Only loop (c) contributes to the Casimir force, since it
``sees'' both surfaces. Here, the loop picks up nonlocal information
about a whole neighborhood, whereas the proximity force approximation
employs only information about local distances indicated by the dashed
line.}
\label{figpierce}
\end{figure} 

Inserting this potential into the worldline formula\re{1.3}, we
encounter the integral
\begin{eqnarray}
I_V[y(t);T,x] :=\int_0^1 dt\, V(x+\sqrt{T}y(t))
&=&\lambda\int_0^1  dt\int_\Sigma d\sigma\, 
  \delta\big(\sqrt{T} y(t)+(x-x_\sigma)\big)
  \nonumber\\ 
&=&\frac{\lambda}{\sqrt{T}} \int_\Sigma d\sigma
  \sum_{\{t_i|\sqrt{T} y(t_i)+x=x_\sigma\}} \frac{1}{|\dot{y}(t_i)|},
  \label{2.2}
\end{eqnarray}
where $\{t_i\}$ is the set of all points where a given scaled unit
loop $\sqrt{T}y(t)$ centered upon $x$ pierces the Casimir surface
$\Sigma$ at $x_\sigma$. If a loop does not pierce the surface (for
given $T$ and $x$), $I_V[y(t)]=0$ for this loop. Of course, there are
also loops that merely touch the Casimir surface but do not pierce
it. For these loops, the inverse velocity $1/|\dot{y}(t_i)|$ diverges
on the surface. But since this divergence occurs in the argument of an
exponential function, these loops remove themselves from the ensemble
average.

As an example, let $\Sigma$ consist of two disconnected surfaces
(bodies), such that $V(x)=V_1(x)+V_2(x)$. For a given propertime $T$,
the Casimir energy density at point $x$ receives contributions only
from those loops which pierces one of the surfaces. The {\em
interaction energy density} defined in Eq.\re{ren2} is even more
restrictive: if a certain loop $y_0(t)$ does not pierce one of the
surfaces, then
$(W_{V_1+V_2}[y_0]-1)-(W_{V_1}[y_0]-1)-(W_{V_2}[y_0]-1)=0$. Therefore,
only those loops which pierce {\em both} surfaces 
contribute to the interaction energy density, as illustrated in
Fig.~\ref{figpierce}.

If the loop $y_0(t)$ does pierce both plates, its contribution to the
energy density is
\begin{equation}
\text{contrib.
of}\,\,\,y_0(t)=1-(\E^{-T\,I_{V_1}}+\E^{-T\,I_{V_2}}-\E^{-T\,I_{V_1+V_2}})
\in (0,1]. \label{2.4}
\end{equation}
From this general consideration, together with the global minus sign in
Eq.\re{1.3}, we learn that Casimir forces between rigid bodies in our
scalar model are always attractive. This statement holds, independent
of the shape of the bodies and the details of the potential (as long
as $V(x)$ is non-negative). 

In the Dirichlet limit, $\lambda\to\infty$, the exponential functions
in Eq.\re{2.4} vanish, and the contribution of a loop is $=1$ if it
pierces both surfaces and $=0$ otherwise. 

\section{Numerical results}
\label{numres} 

\subsection{Parallel Plates}
\label{parplates}

Let us first consider the classic example of a Casimir configuration
consisting of parallel plates separated by a distance $a$ and located
at $z=-a/2$ and $z=a/2$ orthogonal to the $z\equiv x_d$ axis. For
this, Eq.\re{2.1} reduces to
\begin{equation}
V(x)\equiv V(z)=\lambda [\delta(z+a/2) +\delta(z-a/2)]\equiv
V_1+V_2. \label{2.3} 
\end{equation}

\begin{figure}
\begin{center}
\epsfig{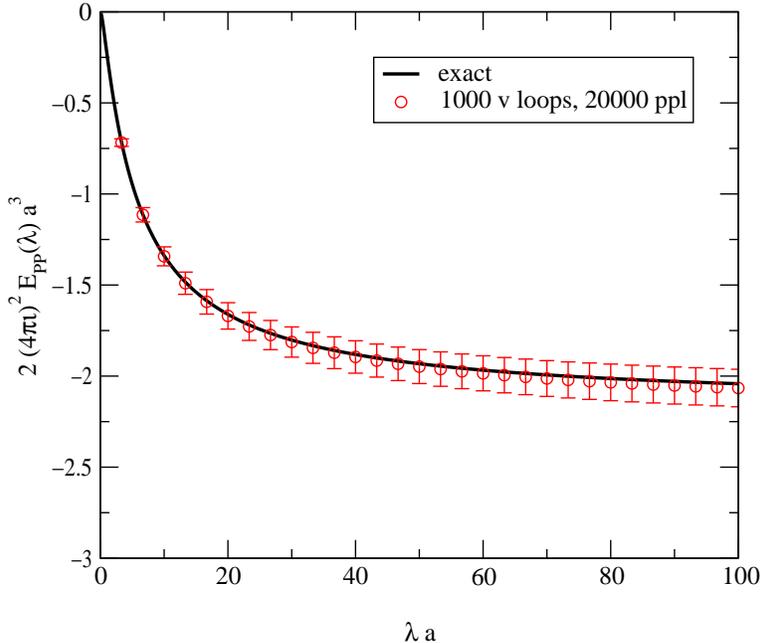}
\end{center}

\vspace{-0.6cm}

\caption{{\bf Parallel plates}: interaction Casimir energy per unit area
for the parallel-plate configuration as a function of the coupling
$\lambda$ (units are set by the plate separation $a$). The numerical
estimate reproduces the exact result for a wide range of couplings
including the Dirichlet limit (cf. Fig.\ref{figppl}).}
\label{eoflambdaPP}
\end{figure} 

In order to test the numerical worldline approach, we compare our
numerical estimates with the analytically known result
\cite{Bordag:cm} of the interaction Casimir energy for arbitrary coupling
$\lambda$ and scalar mass $m$ in units of the plate separation $a$. In
Fig.~\ref{eoflambdaPP}, we study a wide range of couplings and the
approach to the Dirichlet limit, $\lambda a \gg 1$; here, the energy
per unit area tends to
\begin{equation}
\lim_{\lambda a\to \infty} E_{\text{PP}}(\lambda, a) = \frac{1}{2
(4\pi)^2 }\, 
\frac{\pi^4}{45}\, \frac{1}{a^3}\simeq \frac{1}{2 (4\pi)^2 }\, \times
\,2.16\dots\,\times\, \frac{1}{a^3},
\label{CasClas}
\end{equation}
which is the classic Casimir result for a massless scalar
field.\footnote{Here and in the following, we have explicitly
displayed the common propertime prefactors $ 1/[2 (4\pi)^{D/2}]$ for
convenience (see prefactor in Eq.\re{1.3}).} As is visible in
Fig.~\ref{eoflambdaPP}, the agreement is satisfactory even for small
ensembles with $N=20'000$ppl.

\begin{figure}
\begin{center}
\epsfig{figure=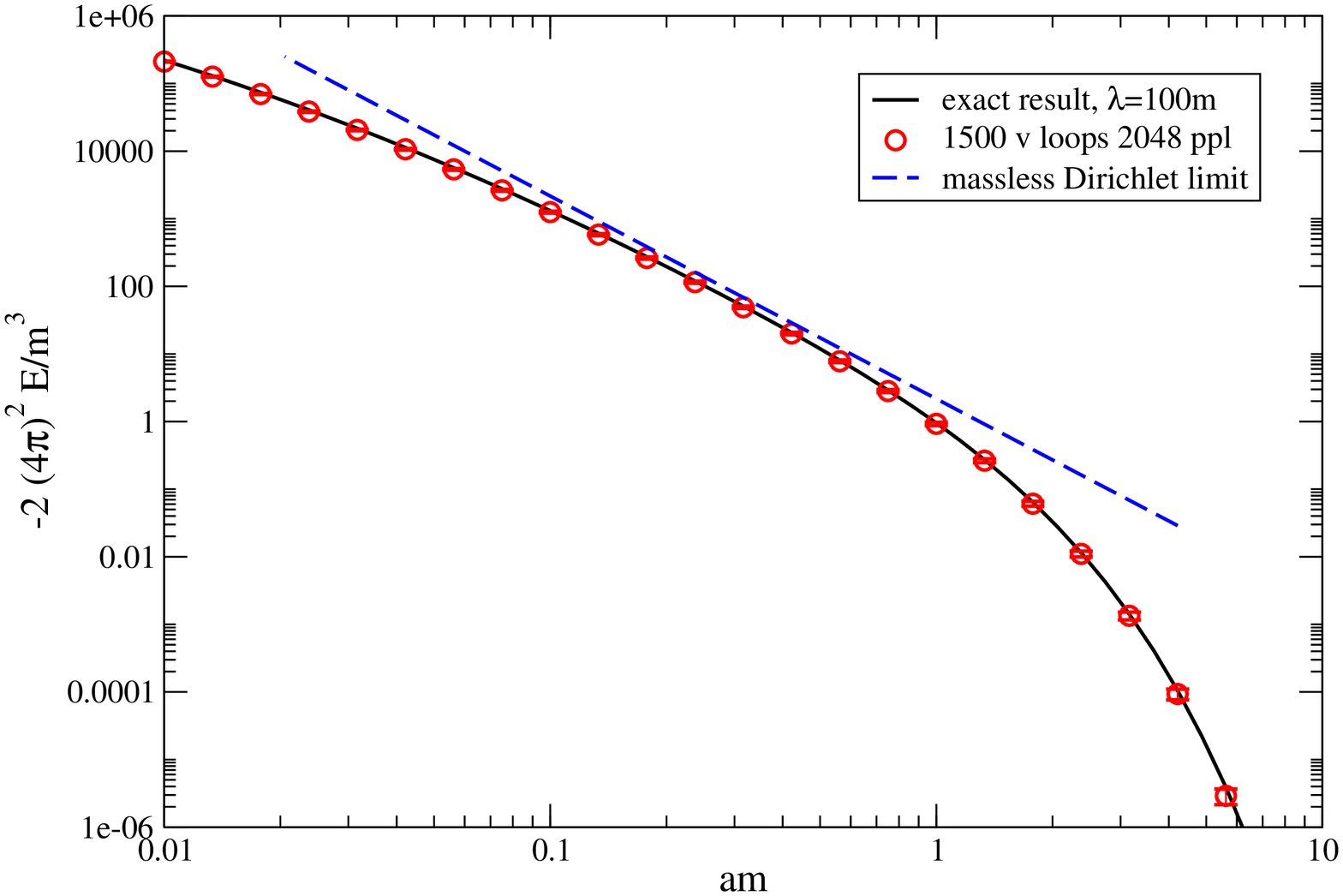,width=12cm}
\end{center}

\vspace{-0.6cm}

\caption{{\bf Parallel plates}: interaction Casimir energy per unit
area for the parallel-plate configuration as a function of the
distance $a$ in units of the mass $m$ for $\lambda =100 m$. The exact
result (solid line) \cite{Bordag:cm} is well reproduced by the
numerical estimate over many orders of magnitude. For intermediate
parameter values, the classic Casimir result (idealized Dirichlet
limit Eq.\re{CasClas}, dashed line) represents a reasonable
approximation.  }
\label{aeps}
\end{figure}

Let us finally discuss the Casimir energy as function of the distance
$a$ of two parallel plates for finite mass $m$ and finite $\lambda $,
in order to explore the strength of the worldline approach in various
parameter ranges.  The result is shown in Fig.~\ref{aeps}.  A finite
value for $\lambda $ simulates a finite plasma frequency.  Hence, for
$a \ll 1/\lambda$ the plates become more transparent for those modes
of the quantum field which fit between the plates. This weakens the
increase of the interaction Casimir energy for decreasing plate
separation, which turns from $\sim1/a^3$ into a $\lambda^2/a$ law
\cite{Bordag:cm}. For $a\gg 1/m$, we observe that the Casimir energy
decreases exponentially with $a$, as expected, since possible
fluctuations are suppressed by the mass gap.  In the intermediate
distance regime, $1/\lambda \ll a < 1/m $, a reasonable approximation
is given by the classic power law $E_{\text{PP}} \sim 1/a^{3}$,
which is familiar from the ideal case $\lambda \rightarrow \infty$, $
m=0$.

\subsection{Sphere above plate}

The Casimir force between a sphere or a spherical lens above a plate
is of utmost importance, because a number of high-precision
measurements have been performed with this experimental
configuration. Let us confine ourselves to the massless case, $m=0$,
in the Dirichlet limit $\lambda\to \infty$; generalizations to other
parameter ranges are straightforward, as in the parallel-plate case. 

\begin{figure}
\begin{center} 
\epsfig{figure=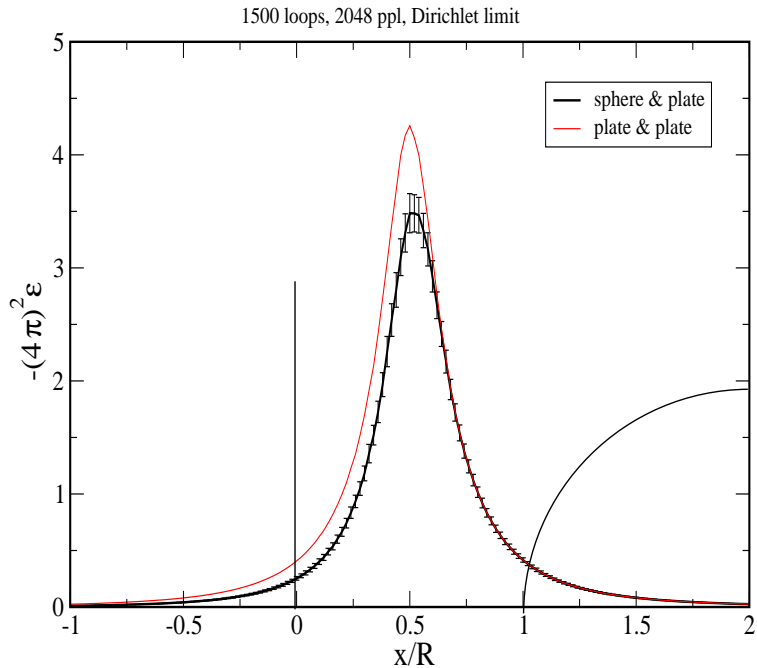,width=10cm}
\end{center}
\caption{{\bf Sphere above Plate}: interaction Casimir energy density
along the symmetry axis ($x$ axis) for the sphere-plate configuration
in comparison to the parallel-plate case. Close to the sphere, the
worldline loops do not ``see'' the curvature; but at larger distances,
curvature effects enter the energy density. For illustration, the
sphere-plate geometry is also sketched (thin black lines).}
\label{sph_den}
\end{figure}

In order to gain some intuition for curvature effects, let us consider
a sphere of radius $R$ the center of which resides over a plate at
distance $a=R$ as an example. The interaction Casimir energy density
along the symmetry axis is shown in Fig.~\ref{sph_den}. For
comparison, the energy density of the case where the sphere is
replaced by a plate is also shown. One observes that the energy
density close to the sphere is well approximated by the energy density
provided by the parallel-plates scenario. This is already at the heart
of the nonlocal nature of the Casimir force and can easily be
understood in the worldline approach.

Recall that the dominant contribution to the interaction Casimir
energy density arises from loops which intersect both surfaces. If
the center of the loop is located close to the sphere, the loops which
intersect both surfaces hardly experience the curvature of the
sphere; this is because loops that are large enough to pierce the
distant plate will also pierce the close-by sphere rather independent
of its radius. By contrast, if the loop center is located close to the
plate, the dominant (large) loops possess intersections with the
sphere at many different points -- not necessarily the closest
point. In this case, the worldline loops ``see'' the curvature of the
sphere that now enters the energy density. 

\begin{figure}
\begin{center} 
\epsfig{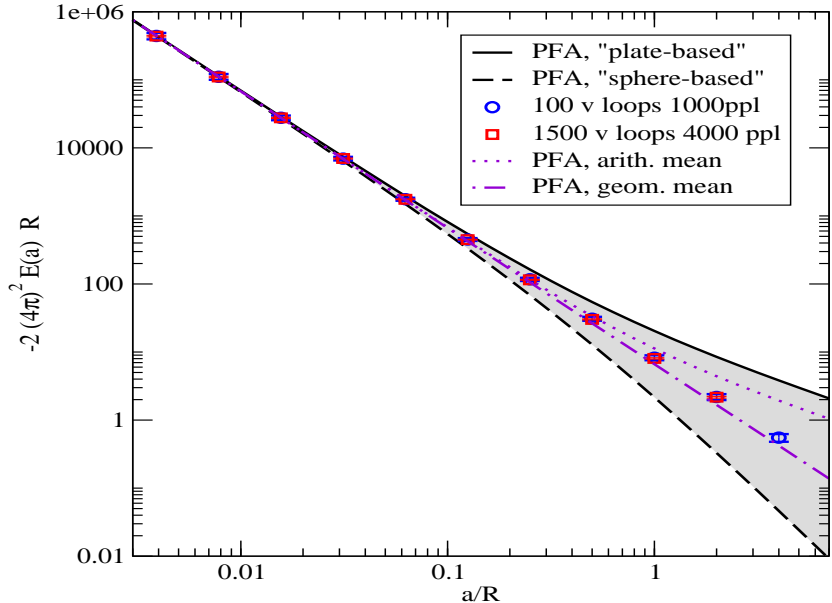}
\end{center}
\caption{{\bf Sphere above Plate}: logarithmic plot of the interaction
Casimir energy for the sphere-plate configuration. For small
separations/large spheres, $a/R\lesssim0.02$, the proximity force
approximation (PFA) approximates the numerical estimate well; but for larger
$a/R$, curvature effects are not properly taken into account. The PFA
becomes ambiguous for larger $a/R$, owing to possible different choices
of the integration domain $S$ in Eq.\re{pft}. A geometric mean
(dotted-dashed line) of $S=S_{\text{plate}}$ and $S=S_{\text{sphere}}$
shows reasonable agreement with the numerical result.}
\label{figSpPl}
\end{figure}

Let us now consider the complete interaction Casimir energy for the
sphere-plate configuration as a function of the sphere-plate distance
$a$ (we express all dimensionful quantities as a function of the
sphere radius $R$). In Fig.~\ref{figSpPl}, we plot our numerical
results in the range $a/R\simeq {\cal O}(0.001\dots 10)$. Since the
energy varies over a wide range of scales, already small loop
ensembles with rather large errors suffice for a satisfactory
estimate (the error bars of an ensemble of 1500 v loops with 4000 ppl
cannot be resolved in Fig.~\ref{figSpPl}). 

\begin{figure}
\begin{center} 
\epsfig{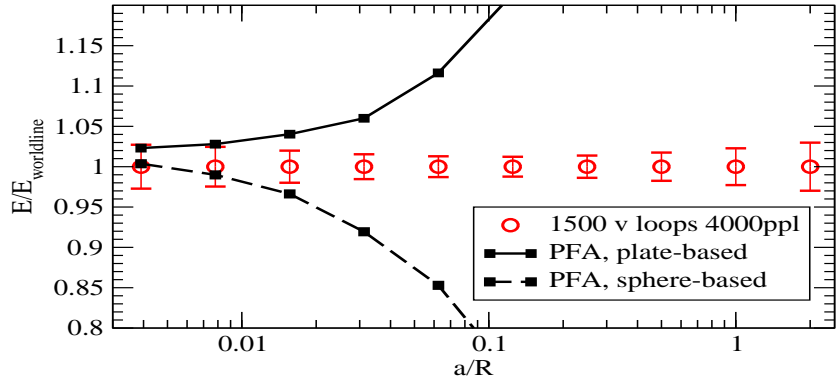}
\end{center}
\caption{{\bf Sphere above Plate}: interaction Casimir energies
normalized to the numerical result (further conventions as in
Fig.~\ref{figSpPl}). For $a/R\gtrsim0.02$, the fluctuation-induced
curvature effects occur at the percent level.}
\label{fignorm}
\end{figure}

Let us compare our numerical estimate with the proximity force
approximation (PFA): using the plate surface as the integration domain
in Eq.\re{pft}, $S=S_{\text{plate}}$, we obtain the solid line in
Fig.~\ref{figSpPl} (PFA, plate-based), corresponding to a
``no-curvature'' approximation. As expected, the PFA approximation
agrees with our numerical result for small distances (large sphere
radius). Sizable deviations from the PFA approximation of the order of
a few percent occur for $a/R\simeq0.02$ and larger. Here, the
curvature-neglecting approximations are clearly no longer valid. This
can be read off from Fig.~\ref{fignorm}, where the resulting
interaction energies are normalized to the numerical result.

In the PFA, we have the freedom to choose alternatively the sphere
surface as the integration domain, $S=S_{\text{sphere}}$. Although
still no curvature-related fluctuation effects enter this
approximation, one may argue that information about the curvature is
accounted for by the fact that the integration domain now is a curved
manifold. Indeed, Fig.~\ref{figSpPl} shows that this ``sphere-based''
PFA approximation deviates from the plate-based PFA in the same
direction as the numerical estimate, but overshoots the latter by
far. It is interesting to observe that the geometric mean, contrary to
the arithmetic mean, of the two different PFA approximations lies
rather close to the numerical estimate; we will comment on this in
more detail in the next section.

\subsection{Cylinder above plate}

In order to study the relation between PFA approximations and the full
numerical estimate a bit further, let us consider a second example of
a cylinder above a plate. Apart from the difference in the third
dimension, all parameters and conventions are as before. 

\begin{figure}
\begin{center} 
\epsfig{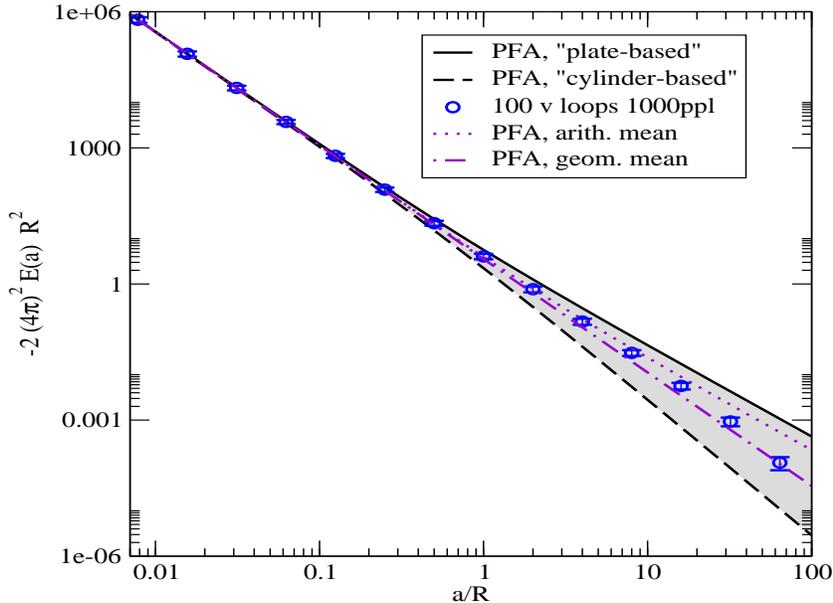}
\end{center}
\caption{{\bf Cylinder above Plate}: logarithmic plot of the interaction
Casimir energy for the sphere-plate configuration
(cf. Fig.~\ref{figSpPl}).}
\label{figCyPl}
\end{figure}

Again, we observe in Fig.~\ref{figCyPl} that the numerical estimate is
well approximated by the PFA for $a/R\lesssim0.02$, but curvature
effects become important for larger distance-to-curvature-radius
ratios. As in the sphere-plate case, the plate-based PFA neglects, but
the cylinder-based PFA over-estimates, the curvature effects for $a/R$
of order one.

Our results seem to suggest that the various possible choices for the
integration domain in the proximity force approximation may give upper and
lower bounds for the correct answer. Indeed, the geometric mean between
the two possible choices for the sphere-plate configuration is rather
close to the numerical estimate (dotted-dashed line in
Figs.~\ref{figSpPl} and \ref{figCyPl}). Similar positive results for
the geometric mean have been found for the two-concentric-cylinder
configuration \cite{Mazzitelli:2002rk} using semiclassical
approximations \cite{schaden} and for a ``chaotic'' geometry
\cite{pft2}.

However, we believe that this ``agreement'' beyond the strict validity
limit of the PFA is accidental. First, detailed inspection reveals
that the geometric mean and the numerical estimate are not fully
compatible within error bars; this is particularly visible in the
cylinder-plate case in Fig.~\ref{figCyPl}. Secondly, there are no
fundamental arguments favoring the geometric mean; by contrast, the
arithmetic mean (as well as the quadratic mean) are not good 
approximations. Thirdly, for even larger separations, $a/R\to \infty$,
it is known that the interaction Casimir energy in the sphere-plate
case behaves as $\sim R^3/a^4$ \cite{163}, whereas even the
sphere-based PFA decreases only with $\sim R^2/a^3$. From the
viewpoint of the worldline, it is obvious anyway that true
fluctuation-induced curvature effects cannot be taken into account by
PFA-like arguments. Nevertheless, the geometric-mean prescription may
yield a reasonable first guess for Casimir forces in a parameter range
beyond the formal validity bounds of the PFA where the expansion
parameter is maximally of order one.

\section{Conclusions}

We have proposed and developed a new method to compute Casimir
energies for arbitrary geometries from first principles in a
systematic manner. The approach is based on perturbative quantum field
theory in the string-inspired worldline formulation which maps field
theoretic problems onto one-dimensional quantum mechanical path
integrals with an evolution in a ``5th coordinate'', the
propertime. These path integrals can easily be performed with
numerical Monte-Carlo techniques. 

Beyond any technical and numerical advantages, we first would like to
stress that the worldline formulation offers an intuitive approach to
the phenomena induced by quantum fluctuations. The geometric
dependence of Casimir forces between rigid bodies, curvature effects
and nonlocalities can already be guessed when thinking in terms of
worldline {\em loop clouds}. 

As to technical advantages, the (usually complicated) analysis of the
fluctuation spectrum and the mode summation are performed at one fell
swoop in the worldline approach. Above all, our algorithm is
completely independent of the details of the Casimir geometry and no
underlying symmetry is required. The algorithm is scalable: if higher
precision is required, only the parameters of the loop ensemble
(points per loop and number of loops) have to be adjusted\footnote{The
numerical computations for this work have been performed on ordinary
desktop PC's. Improvement in precision can be obtained at
comparatively low cost, since the computer resources required increase
only linearly with our loop parameters.}. 

In this work, we have focused on Casimir forces between rigid bodies for
which a computation of the interaction energy suffices; the latter is
free of subtle problems with renormalization. Nevertheless, the
worldline approach is in principle capable of isolating and classifying
divergencies of general Casimir energy calculations, and the
unambiguous program of quantum field theoretic renormalization can be
performed. 

Confining ourselves to a fluctuating real scalar field, we tested our
method using the parallel-plate configuration. New results have been
obtained for the experimentally important sphere-plate configuration:
here we studied the (usually neglected) nonlocal curvature effects
which become sizable for a distance-to-curvature-radius ratio of
$a/R\gtrsim0.02$. Even though the proximity force approximation (PFA)
as standard approximation method cannot correctly account for
fluctuation-induced curvature effects, we found (accidental) agreement
between our numerical estimate and the PFA with a ``geometric-mean
prescription'': the latter implies a geometric mean over the possible
choices of surface integration in Eq.\re{pft}. This geometric mean PFA
might provide for a first guess of the Casimir force for $a/R$ of
order one, but has to be treated with strong reservations.

In this work, we have accepted a number of simplifications, in order
to illustrate our method. Many generalizations to more realistic
systems are straightforward, as discussed in the remainder of this
section: 

\noindent
1) We modeled the Casimir bodies by $\delta$ potentials, mostly
   taking the Dirichlet limit. In fact, this was not a real
   simplification, but numerically even more demanding. Modeling the
   bodies by finite and smooth potential wells requires worldline
   ensembles with a much smaller number of points per loop. The
   $\delta$ potentials represent the ``worst case'' for our
   algorithm, which has nevertheless proved to be applicable. 

\noindent
2) In experimental realizations, effects of finite temperature and
   surface roughness have to be taken into account. Both can be
   implemented in our formalism from first principles. Including
   finite temperature with the Matsubara formalism leads to a
   worldline integral with periodic boundary conditions of the
   worldline loops in Euclidean time direction
   \cite{McKeon:if,Gies:2001tj} which can easily be performed for
   Casimir configurations. The surface roughness can be accounted for
   by adding a characteristic random ``noise'' to the local support of
   the potential. In both cases, the observables can directly be
   computed by our formalism without any kind of perturbative
   expansion.

\noindent
3) For obtaining the Casimir force, our results for the interaction
   energy have to be differentiated with respect to the separation
   parameter. Since numerical differentiation generally leads to 
   accuracy reduction, it is alternatively possible to perform the
   differentiation first analytically; this yields a slightly more
   complicated worldline integrand which can nevertheless be easily
   evaluated without loss of precision. By a similar reasoning, we can
   also obtain the (expectation value of the) energy-momentum tensor, 
   which is frequently at the center of interest in Casimir
   calculations.  For this, we can exploit the fact that the
   energy-momentum tensor can be obtained from the effective action by
   differentiating Eq.\re{1.3} with respect to the metric
   analytically; the resulting worldline integrand can then be put
   into the standard path integral machinery.

\noindent
4) Radiative corrections to the Casimir effect can also be included in
   our method, employing the higher-loop techniques of the worldline
   approach \cite{Schubert:2001he}. We expect these computations to be
   numerically more demanding, since more integrations are necessary,
   but the general approach remains the same. 

\noindent
5) The implementation of finite conductivity corrections is less
   straightforward, since this generally requires a formulation for
   real electromagnetic fluctuations (an extension to complex scalars
   is not sufficient). For this, the starting point can be a field
   theoretic Lagrangian defining a model for the interaction of the
   electromagnetic field with the bodies as suggested, e.g., in
   \cite{itz}. Although these Lagrangians are generally not
   renormalizable, one may expect that the dispersive properties of
   the bodies provide for a physical ultraviolet cutoff (although this
   has to be studied with great care \cite{Barton}).

\section*{Acknowledgment}
We are grateful to W.~Dittrich, M.~Quandt, O.~Schr\"oder and H.~Weigel
for useful information and comments on the manuscript.  We would like
to thank M.~L\"uscher for providing us with the latest
double-precision version of the RANLUX random-number generator.
H.G.~acknowledges financial support by the Deutsche
Forschungsgemeinschaft under contract Gi 328/1-2. L.M.~is supported by
the Deutsche Forschungsgemeinschaft under contract GRK683.

\appendix

%\section{Another worldline prescription for $\delta$ potentials}

\end{document}